# Fundamentals of the Extremely Green, Flexible, and Profitable 5G M2M Ubiquitous Communications for Remote e-Healthcare and other Social e-Applications[1]

Alexander Markhasin

Department of Telecommunication Networks
Siberian Telecommunications University
Novosibirsk, Russia
almar@risp.ru

*Abstract*—The revolutionary trend of the up-to-date medicine can be formulated as wide introduction into basic medicine fields of electronic (e-health) and mobile (m-health) healthcare services and information applications. Unfortunately, all list of qualified m/e-healthcare services can be provided cost-effectively only in urban areas very good covered by broadband 4G/5G wireless communications. Unacceptably high investments are required into deployment of the optic core infrastructure for ubiquitous wide covering of sparsely populated rural, remote, and difficult for access (RRD) areas using the recent (4G) and forthcoming (5G) broadband radio access (RAN) centralized techniques, characterized by short cells ranges, because their profitability boundary exceeds several hundred residents per square km. Furthermore, the unprecedented requirements and new features of the forthcoming Internet of Things (IoT), machine-to-machine (M2M), and many other machine type IT-systems lead to a breakthrough in designing extremely green, flexible, and cost-effective technologies for future 5G wireless systems which will be able to reach in real time the performance extremums, trade-off optimums and fundamental limits. This paper examines the 5G PHY-MAC fundamentals and extremely approaches to creation of the profitable ubiquitous remote e/m-health services and telemedicine as the main innovation technology of popular healthcare and other social e-Applications for RRD territories. Proposed approaches lean on summarizing and develop the results of our previous works on RRD-adapted profitable ubiquitous green 4G/5G wireless multifunctional technologies.

## I. Introduction

As it shows in [1], "The adoption of information and communication technologies (ICT) within the healthcare sector led to the concept of electronic health (e-health)… ICT might be used for a variety of health-related tasks, including communication between patients, doctors, and cares; distant provision of care; remote support to electronic diagnostic medical records; medication adherence control, and so on. ICT in the healthcare sector, if properly used, can significantly contribute to the reduction of management costs and increased efficiency. In this line, e-health substantially reduces the displacement of professionals and patients, globally brings down the cost of medical resources, and makes treatment and health watchfulness more comfortable to patients. All in all, e-health might be considered a revolution in this area. However, a probably more important revolution is taking place due to the use of mobile devices (e.g., smartphones): m-health, which could be defined as the discipline founded on the use of mobile communication devices in medicine or, more specifically, the delivery of healthcare services via mobile communication devices." It ensures a new breakthrough opportunity for provision m/e-Health services in high-quality, ubiquitous, immediately and permanent manner [2-5].

However, all set of qualified smart m/e-health services can be provided only in high urbanized areas with well covered optic infrastructures, because the 4G/5G radio channel ranges (Fig. 1) does not exceed 2-3 km [2-4]. As it follows from a look on the smart cities' infrastructure [2, Fig. 2], the qualified m/e-health services are "tied on a short leash" to optic lines and "asphalt", although they may be the most in demand and effective in rural RRD areas. Unacceptably high investments are required into deployment of the core infrastructure for a ubiquitous wide covering of the RRD areas by the traditional, centralized, architecture of the mobile wireless broadband communications 4G LTE/WiMAX [3,4]. In this paper, we offer the extremely effective green, flexible, low-cost, and radically distributed 5G communications [6,7] as a perspective approach for ubiquitous profitable RRD-covering by smart remote e-Healthcare and other social m/e-Applications.

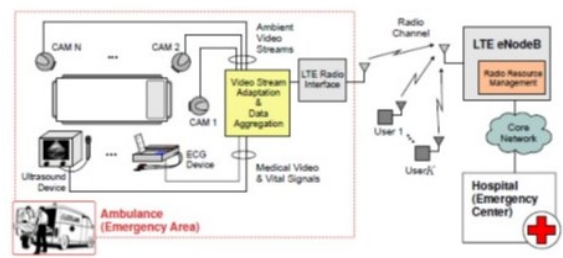

Fig. 1. The m-Health architecture for emergency scenarios [3]

## II. Background of estremely Green and Flexyble 5G

The unprecedented requirements and machine control functions responsible for the forthcoming Internet of Things (IoT), machine-to-machine (M2M), smart e-Healthcare, and many other machine type IT-systems lead a breakthrough jump to design extremely intensive technologies for future 5G wireless systems capable the performance extremes, trade-off

---

optimums and fundamental limits to reach in real time [8,9]. Recently, many 5G extremely intensive solutions which are suitable for well urbanized areas mainly are proposed [2,9,10]. The extremely green and soft themes are proclaimed as 5G two universal major China Mobile themes [11]. We always considered the PHY-MAC's extremal control techniques as very fruitful approach, especially for weakly urbanized areas [6,12-14]. Now, we consider the extremely effective spectral and energy-saving i.e., green, PHYs technology [6,15] and extremely flexible PHY-MACs multifunctional control techniques [7,12-14] as key mission-critical ubiquitous profitable approaches covering by 5G communications of the sparsely populated RRD [16] territories'.

In this paper, we generalize and develop the results researches of the fundamental physical extremums [6,15,17] and information-theoretic limits [12,14], focusing them on the 5G extremely performance issues [6] and PHY-MAC multifunctional optimal control [12, 13] techniques which able to implementing "on-the-fly" the limits closing and invariant criteria optimization/trade-off. The fundamental physical throughput limits and extremums of the energy, power, spectral efficiency invariant criteria [6,15] and limit of the $m$-ary orthogonal signals' interference [17] are described in Section III. In Section IV, the fundamental Shannon borders of the distributed wireless MAC protocols are derived as the maximally reachable throughput capacity, or supreme, and the minimally reachable overhead, or infimum, of the medium access control protocol in practical conditions of multifunctional and multiservice wireless networks [14]. In Section V, are presented our concept of ubiquitous IoT/M2M/H2H green rural 5G communications for remote e/m-Healthcare and other socially significant e-Applications].

## III. GREEN 5G PHY FUNDAMENTAL LIMIT AND EXTREMUMS

### A. The Spectral and Energy Efficiency Invariant Criteria

Usually [9,11,18] and other, the spectral (SE) and energy (EE) efficiency criteria are expressed through the Shannon's capacity of the continuous channels with additive white Gaussian noise (AWGN) [19]

$$C = \Delta F_s \log_2(1 + P_s / P_n), \quad (1)$$

where $\Delta F_s$ is bandwidth, $P_s$ – signal power, $P_n$ – noise power, $P_n = \Delta F_s N_0$, $N_0$ – signal-sided spectral power noise density. The channel's output, or receiver input, powers characteristic $P_s / P_n$ is named as signal-to-noise-ratio (SNR).

However, the continuous channels throughput capacity (1) allows to study only the potential efficiency PHY values depending directly from his three spectral-energy basic parameters. So named, invariant criteria of spectral, power and energy efficiency [6,15,20] allow to solve an optimization or trade-off problem depending from the set of real conditions and parameters of the radio channels, methods of signals coding, formation, modulation, transmitting, receiving, processing, decoding, etc. Two invariant efficiency criteria were firstly introduced for the wireless physical layer with orthogonal spread spectrum $m$-ary signals in [20]. As in [15,20], allow us to introduce an invariant efficiency criterion for modern PHY basing on SINR [21] approaches. The invariant criterion for spectral efficiency (ICSE) was introduced as the digital channel's Shannon capacity per Hertz ((bit/sec)/Hz) [6,15,20]:

$$c_F(m, g, B_s) = C_m(g, B_s)/(B_s/2), \quad (2)$$

where $g$ is channel-side, or receiver input, signal power invariant variable expressed via signal-to-interference plus noise ratio (SINR) [19,20]

$$g^2 = P_s/(P_i + P_n), \quad (3)$$

$P_s$, $P_i$, $P_n$ – signal, interference, and noise powers, respectively; $B_s$ is frequency-time invariant variable named as signal's base, $B_s = 2\Delta F_s T_s$, $T_s$ is $m$-ary signal duration.

Further, $C_m(g, B_s)$ is $m$-ary digital channel's Shannon capacity in bit-per-symbol [15]

$$C_m(g, B_s) = \log_2 m + [1 - p_m(g, B_s)] \log_2[1 - p_m(g, B_s)] + \\ + p_m(g, B_s) \log_2[p_m(g, B_s)/(m-1)], \quad (4)$$

where $p_m(g, B_s)$ is $m$-ary symbol's error probability (SER) [19] defined through invariant variable $h(g, B_s) = g\sqrt{B_s/2}$ expressed, in turn, through receiver's output ratio signal energy per symbol to signal-sided spectral power additional Gaussian interference plus noise density $N_{0_{in}} = N_{0_i} + N_{0_n}$, i.e., symbol energies SINR, or ESINR [15]:

$$h^2 = E_s/N_{0_{in}} = P_s B_s/[2(P_i + P_n)] = g^2 B_s/2 . \quad (5)$$

An invariant criterion for power efficiency (ICPE) was introduced as the signal-to-interference plus noise ratio (SINR) per $m$-ary digital channel's Shannon capacity per Hertz: SINR/[(bit/sec)/Hz] [15]:

$$w(m, g, B_s) = g^2 / c_F(m, g, B_s) . \quad (6)$$

One can express a power efficiency criterion (6) through various measure units: dBm per (bit/sec)/Hz, Watt per (bit/sec)/Hz, and convert to energy efficiency invariant criterion (ICEE) in Joule per (bit/sec)/Hz. Moreover, one can be expressed through invariant criterion for power efficiency (6) the invariant criteria for cover efficiency (ICCE) in Watt/(bit/sec)/Hz/square km, i.e., ICPE per area covering by cell radius $R_c(m, g, B_s)$, and also invariant criterion for investment (cost) efficiency (ICIE) through CAPEX calculated as some invariant function $F_I[w(m, g, B_s)]$ divided into area covering $\pi R_c^2(m, g, B_s)$.

Fig. 2 shows comparison of the invariant spectral efficiency (2) graphs for set OFDM-QAM signals calculated through Shannon $m$-ary digital channel capacity (4) with the potentially achieved Shannon's spectral efficiency (SE) for the continuous AWGN channel. One can see that the proclaimed high values of LTE spectral efficiency can be reached only if SINR more than 20 dBm, i.e., n x 0.1 km cell ranges.

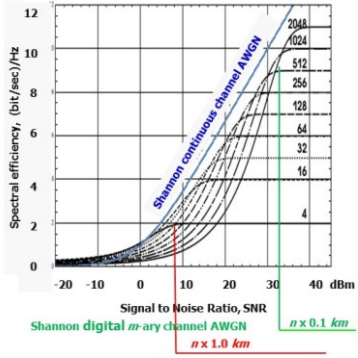

Fig. 2. Comparison of invariant (ICSE) and (SE) spectral efficiency criteria.

### B. The Fundamental Limit and Extremums of the the Spectral and Power Efficiency

The invariant criteria were constructed based on invariant parameters, as introduced above. It is a fact that the value of invariant function $F(x_1, x_2...)$ does not change by substitution instead of every $x_i$ argument's his $x_i^*(x_1, x_2,...)$ invariant maps may be suitable for universal appropriateness research all measures: the information, the power, the covering, and the investment (i.e., profitability) measures. Let us denote by $U$ the set of possible values of the invariant parameters $(m, g, B_s)$. In a specific optimization problem, some invariant variables are free and other parameters are fixed. We denote the set of possible values of the free variables by $V$, $V \in U$. Next we can formulate a set of the optimization problems [15].

#### 1) Power Efficiency Optimization Problem

For an ICPE (6), we can formulate the generally optimization problem

$$w(m, g, B_s) \to \min \quad (7)$$

where free variable belongs to $V$. It is expediently to complete the problem (7) with constraints on the least permissible value of ICSEs

$$c_F(m, g, B_s) \geq [c_F]_{\inf} \quad (8)$$

and, possibly, constraints on the greatest permissible values of cover efficiency ICCE and investment efficiency ICIE, i.e. profitability.

Fig. 3 shows the example of the numerical analysis of ICPE optimization problem. Relying on the analysis graphs of the presented on Fig. 3 we can formulate a fundamental power-consumption

**Statement 1**: The minimal specific power consumption (6) $w_{\min}(m, g, B_s)$ per (bit/sec)/Hz for fixed alphabet size $m$, and free $g$ and $B_s$ for Gaussian both the noise and the interference is an universal power constant which depends neither on the signal base $B_s$, nor from SINR (3).

Let $w_{\min}^*(m, g^*, B_s^*)$ is some infimum point on graph of Fig. 3 which was expressed in SINR-per-(bit/Hz)/sec. We can express this infimum value in Joule-per-(bit/Hz)/sec using the invariant relationship

$$w_{Jc}^*(m, g^*, B_s^*) = w_{\min}^*(m, g^*, B_s^*) \times N_{0_{in}}^* B_s^* / 2, \quad (9)$$

where $N_{0_{in}}$ is the realized in minimum point value of Gaussian both the interference and the signal-sided noise spectral power density as in (5), $N_{0_{in}} = N_{0_i} + N_{0_n}$, in Watt-per-Hertz. Moreover, we can express this minimum value in Joule-per-bit

$$w_{Jb}^*(m, g^*, B_s^*) = w_{Jc}^*(m, g^*, B_s^*) \times B_s^* / 2 \quad (10)$$

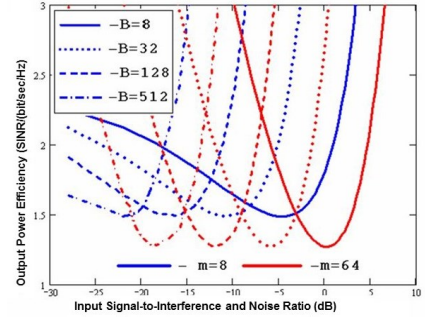

Fig. 3. The powr efficiency optimization's graphs [15].

#### 2) Spectral Efficiency Optimization Problem

For an ICSE (2), we can formulate the generally optimization problem [15]:

$$c_F(m, g, B_s) \to \max \quad (11)$$

with respect to free variables belonging to the subset $V$, $V \in U$. The problem (11) is expediently to complete with constraints on the least permissible value of ICPE criterion

$$w(m, g, B_s) = \mathrm{const}(m) \equiv w_{\inf}(m, g^*, B_s^*) + o(w), \quad (12)$$

where $o(w)$ is small Landau symbol.

The equations (11) and (12) defined the fundamental extremum of the invariant power efficiency ICPE (6) as in

**Statement 2:** The fundamental local maximum, or conditional supreme, of the invariant spectral efficiency (2) under the condition minimal power consumption, or conditional infimum (12), equals solution of the problem (11).

Fig. 4 shows three subsets of ICSE spectral efficiency optimization's graphs accordingly to three fixed values SINRs invariant variable (g = 0,5/1,0/2,0 and different signal alphabet size m with dependency on signal base variations.

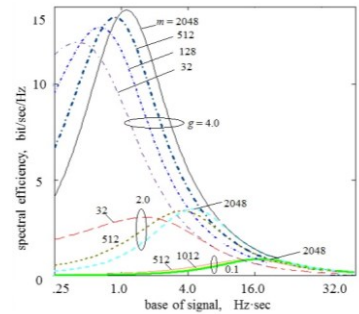

Fig. 4. The spectral efficiency optimization's graphs [6].

In fact, the given series SINR express the changes of channels quality from very poor up to average. Observing the numerical optimization graphs accordingly to the given SINR series and correlating the signal's complexity with signal base values presented on Fig. 4 graphs, we can state following fundamental ICSE statements:

**Statement 3:** Optimal signals accordingly to the spectral efficiency criterion (2) should be as more complicated, as quality SINR of the channel is worse, and, on the contrary, these signals should be easier, when quality of the channel SINR is better];

**Statement 4:** The graphs extremum points represent the fundamental local maximum, or conditional supreme, of the invariant spectral efficiency (2) for a set of SINR under the condition minimal power consumption, or conditional infimum (12).

The above formulated statements lead to optimal extremely green strategies of minimal power consumption and energy saving both the ultra-dense urban (11)-(12) and also the ultra-covering, or extremely cost-effective rural (7)-(8) optimization problems.

### C. Fundamental Limits of m-ary Signals Interference

As shown in [17], the errors of inaccurate fulfillment of conditions of mutual signals orthogonality inevitably generate the intra-cell and inter-cell interference that defines available value of the ratio SINR. The SINR value, in turn, limits the capacity of cellular cell. In [17], the advanced calculation method of the CDMA networks' capacity is offered, which allows to consider the dependences "SINR versus not strict orthogonality errors" directly through the orthogonal signals autocorrelation (ACF) and cross-correlation (CCF) functions.

Let we define the autocorrelation functions ACF and CCF of an $m$-ary orthogonal CDMA signal set as [17]

$$K_{ij}(E) = \int_{t+\varepsilon t}^{t+\varepsilon t+T+\varepsilon T} S_i(t) \cdot S_j(t,E) dt = \begin{cases} 1(E) \leq 1, & \text{if } i = j, \\ 0(E) \geq 0, & \text{if } i \neq j, \end{cases} \quad (13)$$

$$i, j \in \{0,1,...,m-1\},$$

where $E = (\varepsilon_A, \varepsilon_t, \varepsilon_T, \varepsilon_\varpi, \varepsilon_\varphi)$ is vector of the signals' orthogonality errors: signal amplitude $\varepsilon_A$, delays (synchronization) $\varepsilon_t$, duration $\varepsilon_T$, frequencies $\varepsilon_\varpi$ and phases $\varepsilon_\varphi$, $m$ – alphabet volume of orthogonal signals. Further, we can express CDMA system SINR follow [17]:

$$g^2(E) = \frac{[M[K_{ix}^2(E)|_{i=x, i \in I}]]^{1/2} / T}{\sum_{j \in I}[M[K_{jx}^2(E)|_{j \neq i=x}]]^{1/2} / T + \sum_{J \in G \setminus I}\sum_{j \in J}[M[K_{jx}^2(E)|_{x \in I}]]^{1/2} / T + e_{t^\circ}^2}, \quad (14)$$

$$G = \{A, B, ..., I, J, K, ...\}, \quad i, j, x \in \{0, 1, ..., m-1\},$$

where $G$ – set of cells, $m$ – alphabet volume of the orthogonal Walsh sequences or M-subsequences, $K_{ix}^2(E)|_{i=x}$, $K_{jx}^2(E)|_{j \neq x=i}$ accordingly the ACF square and CCF square of $I^{th}$ cell signals, $K_{jx}^2(E)|_{j \in J \in G \setminus I, x \in I}$ – CCF square for other cells, $M[\cdot]$ – expectation, $T$ – signal duration, $e_{t^\circ}^2$ – power of the thermal noise.

Following statements concerning the fundamental limits for interference power are proved [17]:

**Statement 5:** If the errors E caused of the not-strictly orthogonality of the intra-cell $m$-ary orthogonal signals ensembles can be reduce as it's wished, then the power P(E) intracell interference can asymptotically decrease up to as much as small values:

$$\lim_{E \to 0} P_{\text{intra-cell}}(E) = \lim_{E \to 0} \sum_{j=0, j \neq i}^{n-1} \frac{1}{T} \sqrt{M[K_{ji}^2(t, E_j)]} = 0, \quad (15)$$

where $K_{ji}^2(t, E_j)$ is intra-cell cross-correlation function.

**Statement 6:** If the errors E caused of the not-strictly orthogonality of the inter-cell $m$-ary orthogonal signals ensembles can be reduce as it's wished, then the power P(E) inter-cell interference can be asymptotically decrease to small values of an order Landau Big Symbol $0(M[a], 1/\sqrt{n})$:

$$\lim_{E \to 0} P_{\text{inter-cell}} = \lim_{E \to 0} \sum_{J \in G \setminus I} \sum_{J \in G \setminus I} \sum_{j_J}^{n-1} M[a_{j_J}]/T \times$$

$$\times \sqrt{M[K_{j,i}^2(t, E_{j_J})|_{j_J \in J, i \in I}]} = 0(M[a], 1\sqrt{n}), \quad (16)$$

where $K_{j,i}^2(t, E_{j_J})$ is inter-cell cross-correlation function, $M[a]$ is the weighted average of the spaces path loss indexes $a_{j_J}$, $n$ – degree of the generating M-sequence polynomial. Fig. 5 explains impact of the reduction of the signals' orthogonality errors. It is showed, that it is possible to raise many times the networks capacity by reduction of the signals' orthogonality errors.

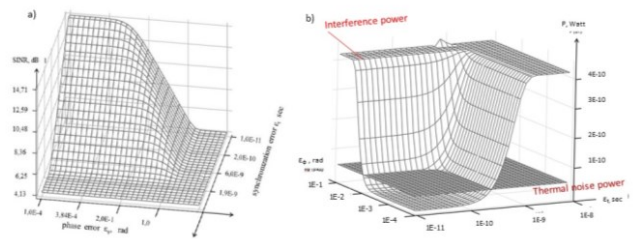

Fig. 5. 3D power graphs versus standard deviations of the synchronization and phase errors, thermal noise -113.101 dB: a) SINR versus errors, b) interference/thermal powers versus errors [17].

## IV. FUNDAMENTAL LIMITS OF DISTRIBUTED 5G MAC

The reachable throughput for various MAC protocols depends on ensuring the "MAC collective intellect" containing a plenitude of information about the real-time state of the multiple access processes in geographically distributed queues, and on the normalized overhead for provisioning QoS. What is the minimum reachable, or infimum, MAC overhead? And what is the potential reachable maximum throughput, or fundamental limit of potential capacity, of the ideal MAC protocol? It is reasonable to find the MAC overhead infimum as the Shannon entropy of the distributed multiple access processes based on the Markov models of distributed queues, and to find the potential capacity of MAC protocols as a function of the overhead infimum [14].

As in [14], let us define the *real throughput capacity* for real MAC protocol specified by real structural specifications and system parameters $\Gamma$, and by real medium conditions $\Psi$ including presence of errors as

$$C_{\Gamma,\Psi} = \max_{\{G \in F_G\}} S_{\Gamma,\Psi}(G), \quad (17)$$

where $F_G$ is the field of the possible values of input traffic intensity $G$. As in [14], we define the *MAC throughput fundamental limit* as supremum of the real throughput (17) on the set $F_\Gamma$ of MAC protocol's possible structural specifications and system parameters by given medium conditions $\Psi$, i.e., as potential capacity,

$$C_\Psi^{sup} = \sup_{\{G \in F_G, \Gamma \in F_\Gamma\}} S_{\Gamma,\Psi}(G) = M[\tau]/(M[\tau]+\delta_\Psi^{inf}) = 1/(1+v_\Psi^{inf}) \quad (18)$$

where $\delta_\Psi^{inf}$ is the potential reachable minimum, or infimum, of the time resource overhead for medium access control per data unit/ packet by duration $M[\tau]$,

$$\delta_\Psi^{inf} = \inf_{\{\Gamma \in F_\Gamma\}} \delta_{\Gamma,\Psi}, \quad (19)$$

is the normalized value of the infimum of overhead (19) according $M[\tau]$.

The fundamental Shannon bounds of the distributed wireless MAC protocols in practical conditions of multiservice wireless networks including the presence of errors, packets priority, various QoS/QoE and queueing models were derived [14], in particular –

*MAC Capacity Theorem:* If the distributed TDMA system is described by the model of equivalent centralized queue $\Gamma_0$ by $\Psi$ medium conditions, the $G$ – upload and $S_{\Gamma_0,\Psi}(G)$ – output traffic intensity, the $m$-ary symbol error rate (SER) $p_{m,\Psi}$ and the $m$-ary channel Shannon throughput efficiency $s_{m,\Psi}$, then the potential throughput capacity of the ideal distributed MAC protocol in presence of errors is equal to

$$\sup_{\{\Gamma_0 \in F_\Gamma; G \in F_G\}} S_{\Gamma_0,\Psi}(G) = C_\Psi^{sup} = \frac{s_{m,\Psi}}{1+\delta_{\Gamma_0,\Psi_0}^{inf}/BM[\tau]s_{m,\Psi}}, \quad (20)$$

the potential reachable minimum, or infimum, of the time resource overhead on distributed MAC per date unit/packet by zero errors $\Psi_0$ equal to

$$\inf_{\{\Gamma_0 \in F_\Gamma\}} \delta_{\Gamma_0,\Psi_0} = \delta_{\Gamma_0,\Psi_0}^{inf} = \begin{cases} 2+H(\tau), & \text{if } \Gamma_0 \to M/M/1 \quad (21a) \\ 2\frac{n+2^{-n-1}}{n+1}+H(\tau), \to M/M/1/n \quad (21b) \\ 1,854+H(\tau), & \text{if } \Gamma_0 \to M/D/1 \quad (216c) \end{cases}$$

where $B$ is the bit rate, $M[\tau]$ is the mean duration of traffic packets, $H(\tau)$ is the entropy of the packets duration distribution given by the geometric law, the throughput efficiency $s_{m,\Psi}$ is m-ary channel Shannon capacity normalized according Hartley capacity.

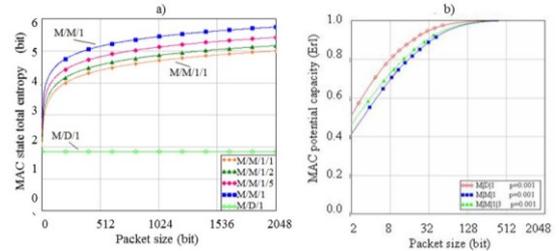

Fig. 6. MACs state entropy, or overhead infimum (a), and MACs throughput supreme(b) versus packets duration laws.

We observe in Fig. 6, that the MAC's total entropy, i.e., overhead infimums (21), depends mainly from the data slots duration law indeterminacy. The M/M/1/* systems family must be characterized by greatest entropy in accordance with its exponential law's greatest indeterminacy. Opposite them, the M/D/1/* systems which are described by deterministic duration law ensure the least entropy, therefore – the greatest MAC potential throughput capacity (20). As proved in [12,14], the adaptive controlled multiple access MAC protocols with deterministic packet size and, hence – the least overheads, allow to reach to a fundamental limit of a MACs throughput capacity which, in turn, as much close to 1,0 as it's wished.

## V. CONCEPT OF UBIQUITOUS IoT/M2M/H2H GREEN RURAL 5G COMMUNICATIONS FOR REMOTE e-APPLICATIONS

As in [6,7,22], the conceptual look of the IP over DVB-2S multifunctional satellite-based fully distributed mesh hybrid 5G networking technology RCS-MFMAC for RRD areas is explained in Fig. 7, RRD Hypercells architecture – further in Fig. 8.

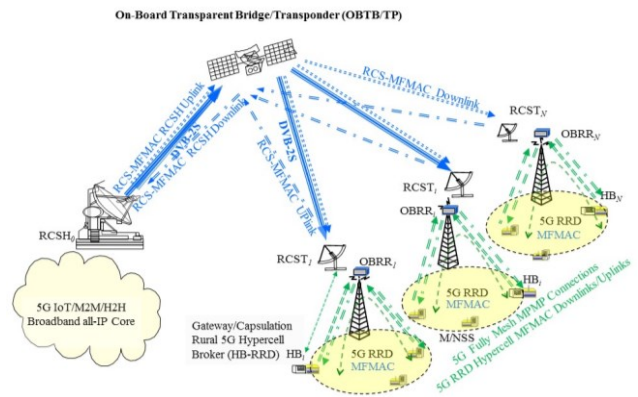

Fig. 7. Device-centric Ubiquitous Green Rural 5G Hybrid Architecture

The hybrid architecture relies on implementation of the QoS-guaranteed multifunctional 5G machine type MAC perfect rural PHY-MAC techniques basing on the developing of the advanced delay-tolerant 5G ATM-like MPMP MFMAC technologies [7,12,16,22] which in turn should be adapted to conditions of the satellite platforms' DVB-2S-RCS [16] VSAT, etc. The main breakthrough drivers for RRD-oriented 5G communications include also a push MFMAC-based next generations of wireless asynchronous transfer mode (ATM/MFMAC), of multi-protocol label switching (MPLS/MFMAC), and of IP over DVD-S/MFMAC integrated networking technologies [16].

## VI. CONCLUSION

In this paper, the extremely effective 5G PHY-MAC fundamentals which open the effective approaches for implementation of the extremely green energy-saving and cost-saving 5G-aimed techniques are proved: i) smarter increase of the SINR through beamforming/ antenna/ orthogonality gain, without rise of the transmitter power; ii) to close "on-the-fly" the fundamental minimum of power consumption ICPE; iii), to implement "on-the-fly" the profitability-power-efficiency-aimed fundamental trade-offs for rural green 5G networks in practical invariant variables notions. The derived results will be useful for design of the extremely effective green PHY and extremal flexible PHY-MAC multifunctional 5G M2M communications for ubiquitous profitable covering by smart remote e-Healthcare and other social significant m/e-Applications on the RRD territories.

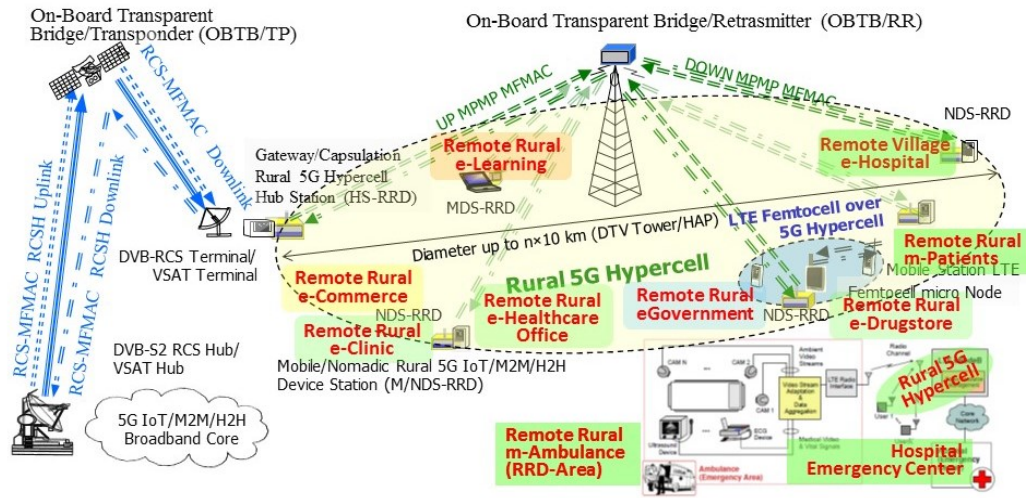

Fig. 8. Rural Green 5G Hypercell Architecture for ubiduitous remote m/e-Healthcare and other other social significant m/e-Applications